\documentclass[aps,prl,twocolumn,superscriptaddress,nofootinbib]{revtex4}
\usepackage{graphicx}
\usepackage{natbib}
\usepackage[breaklinks]{hyperref}
\usepackage{amsmath}
\usepackage{mathtools}
\usepackage[usenames, dvipsnames]{xcolor} 

\begin{document}


\title{Experimental SWAP test of infinite dimensional quantum states}
\author{Chi-Huan Nguyen}
\author{Ko-Wei Tseng}
\author{Gleb Maslennikov\footnote[2]{Present address: NKT Photonics, Bregnerødvej 144, 3460 Birkerød, Denmark}}
\author{H.~C.~J.~Gan}
\affiliation{Centre for Quantum Technologies, National University of Singapore, 3 Science Dr 2, 117543, Singapore}
\author{Dzmitry Matsukevich}
\affiliation{Centre for Quantum Technologies, National University of Singapore, 3 Science Dr 2, 117543, Singapore}
\affiliation{Department of Physics, National University of Singapore, 2 Science Dr 3, 117551, Singapore}

\begin{abstract}  
Efficient overlap estimation of high-dimensional quantum states is an important task in quantum information and a core element in computational speedups of quantum machine learning. Here we experimentally demonstrate the SWAP test that measures the overlap of two motional states in a system of trapped $^{171}\mathrm{Yb}^+$ ions. 
To illustrate the versatility of our implementation, we report the overlap measurement of a variety of quantum states: Fock states, coherent states, squeezed vacuum states, and cat states.
We highlight applications of the SWAP test by measuring the purity of mixed states.
Our results enable quantum information processing with high dimensional quantum states.
\end{abstract}
\maketitle

Estimation of the overlap $\textrm{Tr}(\rho_1 \rho_2)$ between two quantum states $\rho_1$ and $\rho_2$ is a frequently encountered task in quantum information processing~\cite{ArturSwap2002}, quantum computations~\cite{LauCVgates2016}, quantum machine learning~\cite{Biamonte2018} and quantum neural networks~\cite{Zhao2019}. It enables quantum fingerprinting~\cite{qfingerprintPRL2001, SandersFingerprint2005}  which determines whether two $n$-bit strings are identical in $\mathcal{O}(\log_2(n))$ steps. Used in some implementations of variational quantum eigensolvers~\cite{Higgott2019VQE}, it is also an integral part of the quantum support vector machines~\cite{RebentrostPRL2014}. Several important properties of quantum systems such as the fidelity, the purity, the Hilbert-Schmidt distance, and the Renyi entropy can be derived from the overlap estimation~\cite{nielsen_chuang_2010,RadimSWAP2002,IslamNature2015,MonroeSWAP2018}.

A brute force approach to estimating the overlap is first reconstructing the density matrices of quantum states $\rho_1$ and $\rho_2$ using state tomography, followed by calculating $\textrm{Tr}(\rho_1 \rho_2)$~\cite{nielsen_chuang_2010}. As the dimension of the Hilbert space increases, the number of operations required for full state tomography grows exponentially, making this technique computationally costly~\cite{Cotler2020}. 
Alternatively, if the unitary transformation $U_1(x)$ to prepare the quantum state $|x_1\rangle$ starting from the initial state $U_1(x_1)|x_{\textrm{init}}\rangle$ is known, one can measure the overlap $|\langle x_2|x_1\rangle|^2$ by applying the inverse unitary transformation $U_1^{\dagger}(x_1)$ to the other state $|x_2\rangle$ and measure the probability that the final state is $|x_{\textrm{init}}\rangle$~\cite{Ouyang2020a,johri2020nearest,peters2021machine}.

A universal and constant-depth circuit measuring the overlap between quantum states, that are not known \textit{a priori}, would provide quantum advantages for several quantum machine learning algorithms~\cite{Rebentrost2014,Lau2017,Schuld2019}.
This can be realized using the SWAP test that relies on the controlled SWAP gate~\cite{qfingerprintPRL2001} and have been implemented with qubits on various platforms such as photons~\cite{Cai2015,Travnicek2019}, trapped ions~\cite{MonroeSWAP2018}, and superconducting qubits~\cite{Cincio_2018}.
On the other hand, computation can also be carried out with continuous variables (CV)~\cite{Braunstein2005} or a hybrid approach~\cite{Andersen2015} that utilizes both discrete and continuous degrees of freedom of a quantum system. This approach which was pioneered in optics~\cite{Furusawa2008} provides hardware-efficiency, as the same apparatus can encode and process more information using the states of harmonic oscillators if they are inherently available. 
For states of bosonic modes, the SWAP test~\cite{RadimSWAP2002,SWAP_HOM_2013} was first demonstrated using dispersive coupling between a transmon and superconducting cavities~\cite{Gao2018,GaoESwap2019}. The motional modes of trapped ions are another well-controlled bosonic system for exploring quantum information processing over infinite dimensions~\cite{Ortiz-Gutierrez2017,Leibfried2003,Um2016,Kienzler2017,Toyoda2015}.
In this Letter, we demonstrate the SWAP test of high dimensional quantum states encoded in the motional modes of trapped ions.

\begin{figure} [ht!]
\centering
   \includegraphics[width=\columnwidth]{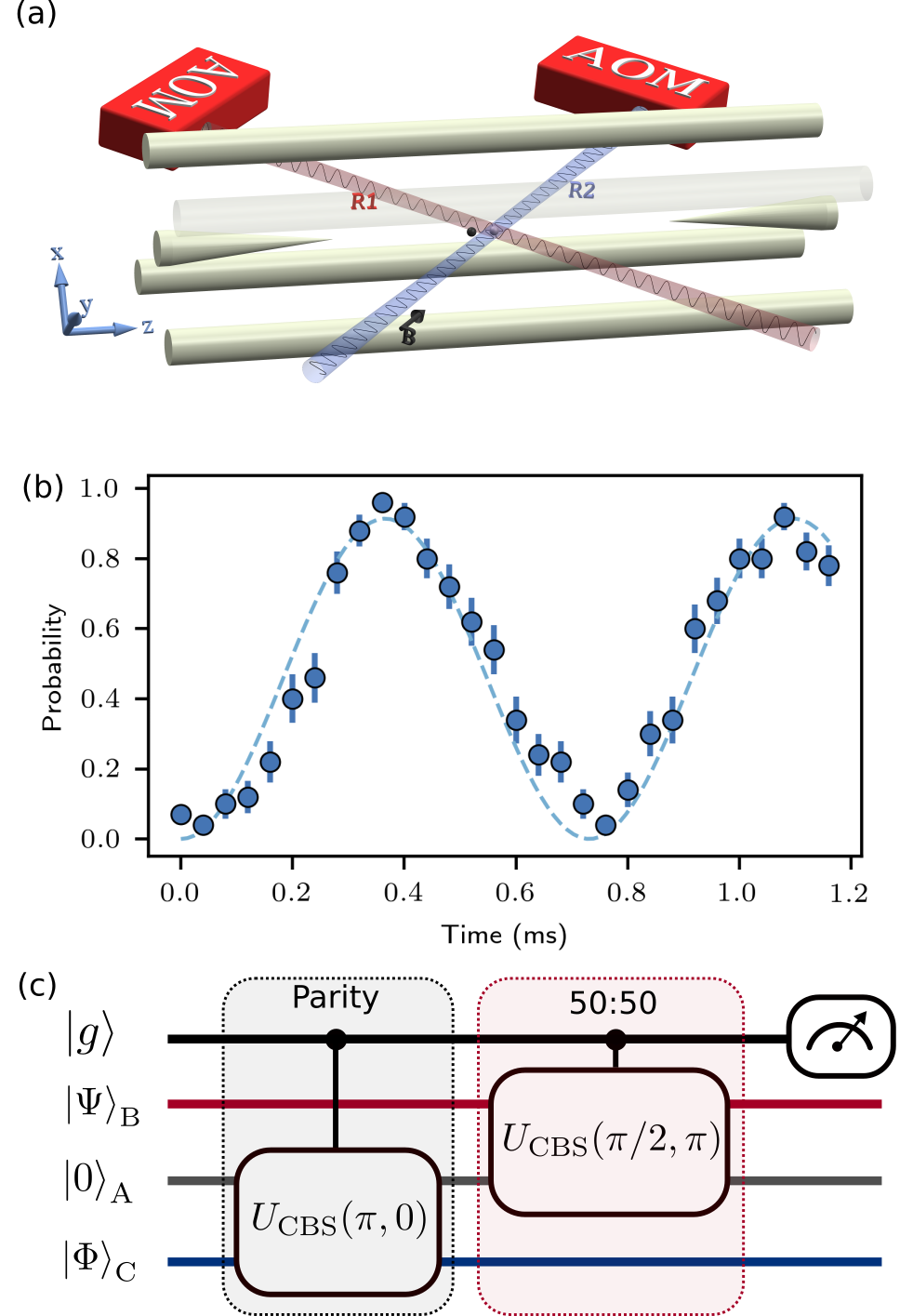}
  \caption{\label{fig:figure1} 
    (a) Experimental setup. A linear rf-Paul trap confines two $^{171}\mathrm{Yb}^+$ ions. Raman beams represented by blue and red beams form a running optical lattice at the ions. The beat note frequency and amplitude of the lattice are controlled by rf signals sent to acousto-optical modulators (AOMs) in each Raman beam.
    (b) Measured $P_e$ as a function of the spin-dependent beam splitter gate duration. Error bars correspond to statistical uncertainty and denote one standard error of the mean.
    The dashed line shows the fit to the sine function to extract the coupling strength $\Omega_0$ of the beam splitter gate.
    (c) SWAP test circuit to measure the overlap of motional states prepared in mode $B$ and $C$.
    }
\end{figure}
In our experiment, two $^{171}\mathrm{Yb}^+$ ions are confined in a linear rf-Paul trap. 
The secular trapping frequencies are $(\omega_x,\omega_y,\omega_z) = 2 \pi \times (0.945, 1.274, 0.519)$\,MHz and actively stabilized with a drift of less than $100$\,Hz/ hour.
We encode the information into the out-of-phase motion in the $y$ direction (mode $A$), out-of-phase and in-phase motion in the $x$ direction (mode $B$ and $C$ respectively) (Fig~\ref{fig:figure1}a). The motional mode frequencies are $(\omega_{A},\omega_{B},\omega_{C})=2\pi \times (0.782, 1.159, 1.274)$\,MHz.
To mediate the interaction between motional modes, two hyperfine levels of one ion are used as qubits, denoted as  $|g\rangle = |^{2}S_{1/2},F=0,m_{F}=0\rangle$ and $|e\rangle = |^{2}S_{1/2},F=1,m_{F}=0\rangle$.
Here, $F$ and $m_{F}$ denote the quantum numbers associated with the total atomic angular momentum and its projection along the quantization axis defined by an applied magnetic field of 5.2\,G, respectively.
The transition frequency of internal states is $\omega_0 /2\pi=12.643$\,GHz.
Standard optical pumping and resonance fluorescence state detection techniques are used to initialize and detect the qubits~\cite{Olmschenk2007}.
The other ion is optically pumped to the metastable state $^{2}F_{7/2}$ by driving the $^{2}D_{3/2} \rightarrow$ $3[1/2]^{\circ}_{3/2}$ transition at $398.98$\,nm. This ``dark'' ion does not interact with the optical fields during the experiment. 

We employ a stimulated Raman process to coherently control the qubit as well as the motion states of ions.
A frequency-doubled, mode-locked Ti:sapphire laser (pulse duration 3 ps, repetition rate 76 MHz) produces a pair of optical beams that are responsible for the Raman transition via the internal states $|g\rangle$ and $|e\rangle$~\cite{Hayes_freqcomb}.
The detuning and the pulse shape of the Raman beams are controlled by acousto-optical modulators (AOMs) (Fig.~\ref{fig:figure1}a).
The single qubit rotation $R(\theta) = \cos (\theta /2)I - i \sin (\theta /2) \sigma_y$ is performed by tuning the Raman beat-note to qubit resonance $\omega_0$.
Here, $\sigma_y=-i|g\rangle \langle e| + i|e\rangle \langle g|$.
Measurement of motional states is done by coupling to the qubit via the blue (bsb) or red (rsb) motional sideband pulses which are described by the time evolution of the Hamiltonians $H_{\textrm{bsb}} = \frac{\hbar \Omega_{\textrm{sb}}}{2}(\sigma^{+}a^{\dagger}+\sigma^{-}a)$ and $H_{\textrm{rsb}} = \frac{\hbar \Omega_{\textrm{sb}}}{2}(\sigma^{-}a^{\dagger}+\sigma^{+}a)$, where $\sigma^{+}=|e\rangle \langle g| $ and $\sigma^{-}=|g\rangle \langle e| $.

The building block for the SWAP test is the spin-dependent beam splitter gate~\cite{JarenCBS2020} $U_{\textrm{CBS}}(\theta,\psi)=\exp[-(i / \hbar) \int_{0}^{T} H_{\textrm{CBS}}(\psi,t)dt)]$ that induces the interaction
\begin{equation}
  \label{eq:CBS_H}
  H_{\textrm{CBS}}(\psi)/\hbar=\frac{\Omega(t)}{2}\sigma_x(a ^{\dagger}b e^{i\psi} +a b ^{\dagger} e^{-i\psi})
  \end{equation}
between two modes of motion conditional on the internal states of the ions denoted as $|g\rangle$ and $|e\rangle$. 
Here, $a (a^{\dagger})$ and $b(b^{\dagger})$ are the annihilation (creation) operator of the motional modes and $\sigma_x=|g\rangle \langle e| + |e\rangle \langle g|$. $\psi$ represents an experimentally tunable phase, $\Omega(t)$ the coupling strength, and $\theta=\int_{0}^{T} \Omega(t) dt$ denotes an effective mixing angle for a gate duration $T$.

We achieve the beam splitter transformation between modes $A$ and $B(C)$ by driving simultaneously two Raman transitions with beat-note frequencies tuned to $\omega_0 \pm |\omega_A-\omega_{B(C)}|$.
To reduce off-resonant coupling to the carrier transition, the actual waveform $\Omega(t)$ that is applied on the AOMs to implement $U_{\textrm{CBS}}$ have smooth rising $\Omega_0 \sin^2{\frac{\pi}{2\tau}t}$ and falling $\Omega_0 \sin^2{\frac{\pi}{2\tau}(T-t)} $ edges with a time constant $\tau=50$\,us, and a constant amplitude $\Omega_0$ in the middle.

The SWAP test is implemented as a combination of two spin-dependent beam splitter gates followed by the measurement of the qubit state as shown in the Fig~\ref{fig:figure1}c. The qubit is initially prepared in the state $|g\rangle = (|+\rangle - |-\rangle)/\sqrt{2}$ where $|\pm\rangle$ are the eigenstates of the $\sigma_x$ operator. After the first gate $U_{\text{CBS}}(\pi, 0)$ the creation operators evolve as $c^{\dagger} \rightarrow (\pm i) a^{\dagger}$,  $a^{\dagger} \rightarrow (\pm i) c^{\dagger}$ conditioned on the qubit being in the state $|\pm\rangle$. If the mode $A$ is initially in the vacuum state and mode $C$ has $n_c$ excitations, this transfers population of the mode $C$ to $A$ and introduces an additional factor $(-1)^{n_c}$, acting as a parity gate for the state $|\Phi\rangle_C$.

The second operator $U_{\textrm{CBS}}(\pi/2, \pi)$ acts as the spin-dependent 50:50 beamsplitter and transforms modes $A$ and $B$ as
$a^{\dagger} \rightarrow (a^{\dagger}     \mp  i b^{\dagger})/\sqrt{2}$, 
$b^{\dagger} \rightarrow (\mp i a^{\dagger} +    b^{\dagger})/\sqrt{2}$, depending on the state of the control qubit. 
The outcome of the qubit measurement cannot be changed by the unitary transformation of only motional modes after the second $U_{\textrm{CBS}}$ gate.
A combination of spin-independent 50:50 beamsplitter gate between mode $A$ and $B$ and a 100:0 beamsplitter gate between mode $A$ and $C$ would return modes $A$ to the vacuum state, and $B$, $C$ to the initial state if the control qubit is in state $|+\rangle$, while transforming modes $B$ and $C$ as $b^{\dagger} \rightarrow c^{\dagger}$, $c^{\dagger} \rightarrow b^{\dagger}$ if the qubit is in the state $|-\rangle$. This is equivalent to the controlled SWAP gate. The overlap of the states in mode $B$ and $C$ is then given by $|_{B}\langle \Psi|\Phi\rangle_{C}|^2=|1-2P_g|$, where $P_g$ is the probability to detect the qubit in state $|g\rangle$~\cite{qfingerprintPRL2001} (see Supplementary Material for a detailed proof). 

 A typical experiment sequence starts with 4\,ms of Doppler cooling followed by 10\,ms of Sisyphus cooling~\cite{Ejtemaee2017}.
The radial motional modes are further sideband cooled to the ground states.
The heating rates are $(\dot{n}_{A},\dot{n}_{B},\dot{n}_{C} )= (0.8, 0.9, 20.2) s^{-1}$.
  \begin{figure} [!ht]
\centering
   \includegraphics[width=\columnwidth]{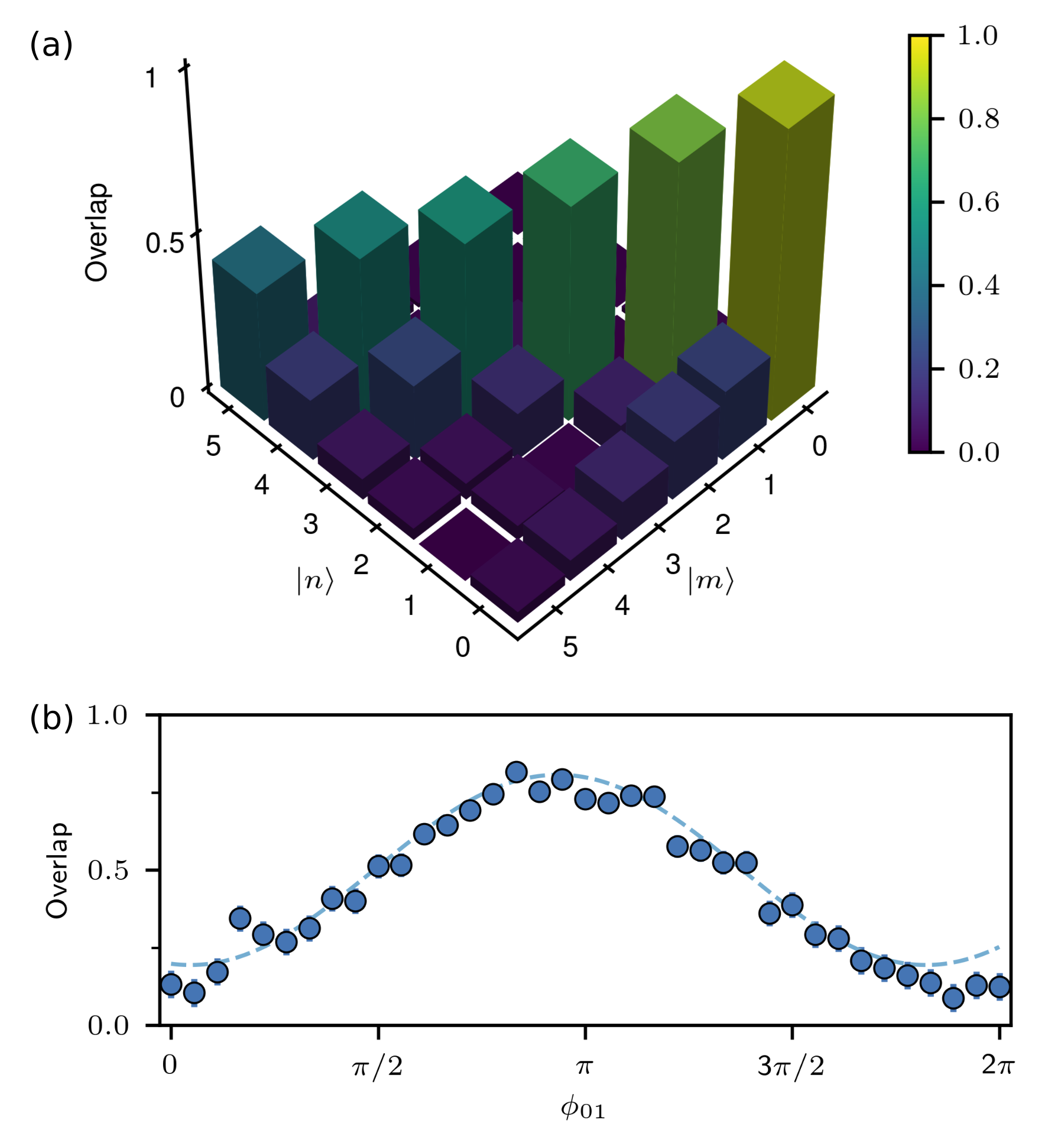}
  \caption{\label{fig:figure2} 
    SWAP test of finite-dimensional quantum states. (a). Measured overlap of Fock states $|\Psi \rangle_{B} = |m \rangle$ and $|\Phi \rangle_C = |n \rangle$
    (b). Overlap of the $|\Psi \rangle_{B}=\frac{1}{\sqrt{2}}(|0\rangle + |1 \rangle)$ and $|\Phi \rangle_{C}=\frac{1}{\sqrt{2}}(|0\rangle -e^{i\phi_{01}}|1\rangle)$ as a function of the relative
    phase difference $\phi_{01}$. Error bars correspond to statistical uncertainty and denote one standard errors of the mean. The solid line fits the data points to a sine function.
    }
  \end{figure}
  \begin{figure*} [ht]
\centering
   \includegraphics[width=\textwidth]{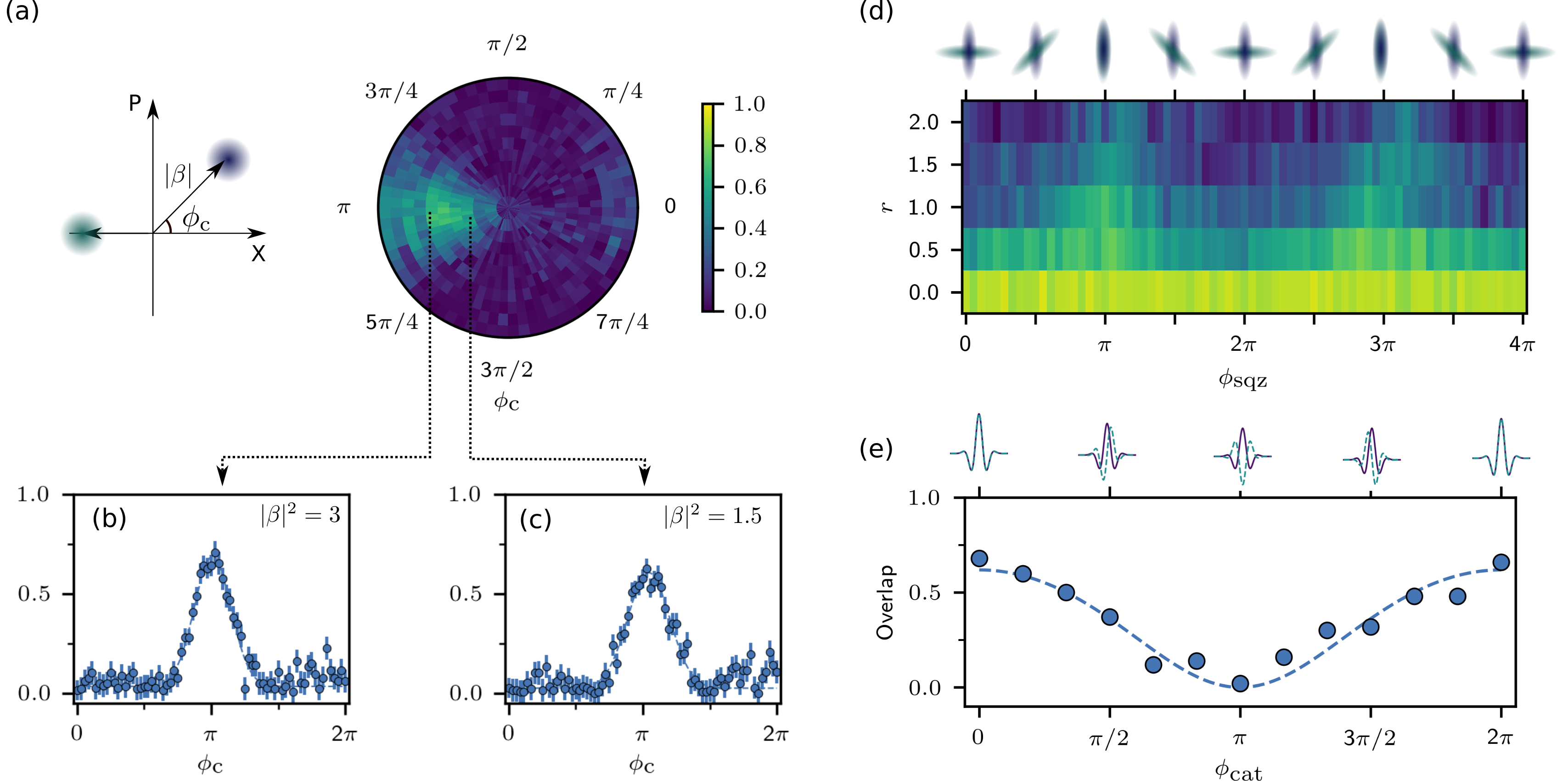}
  \caption{\label{fig:figure3} 
   SWAP test of infinite-dimensional quantum states. (a) Measured overlap of coherent states $|\Psi \rangle_{B} =| e^{i\pi}\alpha \rangle$  and $|\Phi \rangle_{C}=|e^{i\phi_{\textrm{C}}}\beta \rangle$ as a function of the relative displacement angle $\phi_{\textrm{C}}$, for $|\alpha|^2=3$ and $|\beta|^2 \in \{0.0, 0.5, 1.0, 1.5, 2.0, 2.5, 3.0, 3.5, 4.0, 4.5, 5.0\}$.
    1D cuts at displacements (b) $|\beta|^2={3.0}$ and (c) $|\beta|^2={1.5}$.
The dashed lines fit the data to a Gaussian function.
(d) Overlap of the squeezed vacuum states $| \Psi \rangle_B=|re^{i\pi/2} \rangle$ and  $|\Phi_{C}\rangle=|re^{i\phi_{\textrm{sqz}}/2} \rangle$.
Top row shows the illustrated Wigner functions.
(e) Measured overlap of cat states   $|\Psi\rangle_{B} = (|\alpha_1 \rangle + | -\alpha_1 \rangle)/\sqrt{2}$ and
$|\Phi_{C}\rangle = (|\alpha_2 \rangle + e^{i\phi_{\textrm{cat}}}|-\alpha_2 \rangle)/\sqrt{2}$ for  $|\alpha_1|^2 = 0.95(5)$, $ |\alpha_2|^2   = 1.10(6)$ as a function of a variable phase difference $\phi_{\textrm{cat}}$.
Top row shows the imaginary part of Wigner functions of cat states.
The dashed line is a fit to the overlap function of cat states (see text). All error bars correspond to statistical uncertainty and denote one standard error of the mean.
    }
\end{figure*}
We characterize the coupling strength by applying $U_{\textrm{CBS}}$ to the input state $|g\rangle|1\rangle_{A}|1\rangle_{B}$ and measure the probability to detect the internal state in $|e\rangle$, denoted as $P_e$.
Figure \ref{fig:figure1}b shows the measured probability $P_e$ as a function of the gate duration $T$ along with a sinusoidal fit $P_e=P_0 \sin^2{T \Omega_0}$ with fitting parameters $P_0$ and $\Omega_0$.
The coupling strength $\Omega_0/2 \pi \sim 680$\,Hz corresponds to implementing a beam splitter operation $U_{\textrm{CBS}}(\theta = \pi/2)$ in $T \sim 368$\,$\mu$s. A total time required for the SWAP test of $\sim 1.1$\,ms, is independent of the dimension of the input state and is much shorter than the $\sim10$\,ms motional phase coherence time for the superposition of Fock states $(|0\rangle+|1\rangle)/\sqrt{2}$. 

We highlight the versatility of our method by measuring the overlap of a variety of quantum states using the same time sequence and parameters for the SWAP test circuit.
We begin with Fock states $| \Psi \rangle_B=|m\rangle$ and $| \Phi \rangle_C=|n\rangle$, for $n,m \in [0,5]$.
The result shown in Fig.~\ref{fig:figure2}a agrees with the expected outcome of high overlap probability along the diagonal.
We observed an increase in the overlap between $|\Psi \rangle_B=|n+1\rangle$ and $|\Phi \rangle_C=|n\rangle$. Figure \ref{fig:figure2}b shows the result of the SWAP test of superposition states $|\Psi \rangle_B =(| 0 \rangle + |1 \rangle)/\sqrt{2}$ and
$|\Phi \rangle_C =(| 0 \rangle - e^{i\phi_{01}}|1 \rangle)/\sqrt{2}$ as a function of relative phase difference $\phi_{01}$, along with a fit to a sinusoidal function.

We highlight the state-independence of the SWAP test by estimating the overlap of high-dimensional states such as coherent states, squeezed vacuum states, and cats states as shown in Fig.~\ref{fig:figure3}.
We prepared these families of states using spin-dependence forces that results from a running optical lattice formed by the Raman beams~\cite{Ding2014}.
Figure~\ref{fig:figure3}a presents the overlap measurements of coherent states $|\Psi \rangle_{B} =| e^{i\pi}\alpha \rangle$ and $|\Phi \rangle_C =|e^{i\phi_{\textrm{C}}} \beta \rangle$.
As indicated in Fig.~\ref{fig:figure3}b and c, the maximum overlap is observed at the $\phi_{\textrm{C}}=\pi$ (mod $2\pi$) and $|\beta|=|\alpha|$.
We also observe a noticeable overlap at $\phi_{\textrm{C}} \sim 2 \pi$, that can be attributed to the imperfect state preparation.
Figure \ref{fig:figure3}d presents results of SWAP test of squeezed vacuum states  $| \Psi \rangle_B=|re^{i\pi/2} \rangle$ and  $|\Phi \rangle_{C}=|re^{i\phi_{\textrm{sqz}}/2} \rangle$, where $r$ is the squeezing parameter.
The overlap peaks measured as a function of $\phi_{\textrm{sqz}}$ becomes sharper as $r$ increases.
Using the techniques reported in~\cite{Ding2017,Kienzler2016}, we prepare the cat states $|\Psi\rangle_{B} = (|\alpha_1 \rangle + | -\alpha_1 \rangle)/\sqrt{2}$ and
$|\Phi \rangle_{C} = (|\alpha_2 \rangle + e^{i\phi_{\textrm{cat}}}|-\alpha_2 \rangle)/\sqrt{2}$ for $|\alpha_1|^2 = 0.95(5)$ and $ |\alpha_2|^2   = 1.10(6)$  sequentially in mode $B$ and $C$ respectively.
Figure~\ref{fig:figure3}e presents their overlap as a function of relative parity angle  $\phi_{\textrm{cat}}$.
For $|\alpha_1|^2=|\alpha_2|^2 \equiv |\alpha|^2$, the overlap function of the cat states has a form of
\begin{equation}
  \label{eq:cat_overlap}
\gamma_{\textrm{cat}}\frac{(1+\cos{\phi_{\textrm{cat}}})(1+e^{-2|\alpha|^2})}{2(1+\cos{\phi_{\textrm{cat}}}e^{-2|\alpha|^2})}
\end{equation}
where $\gamma_{\textrm{cat}}$ represents the reduction of the overlap due to technical imperfections of our experiment.
From a fit of the overlap to Eq.~\ref{eq:cat_overlap} with two free parameters $G$ and $|\alpha|^2$ (dashed line in Fig.~\ref{fig:figure3}e), we obtain $\gamma_{\textrm{cat}}=0.62(4)$ and the coherent amplitude $|\alpha|^2=0.9(1)$, which agrees with the independently measured values of $|\alpha_1|^2=0.95(5)$ and $|\alpha_2|^2=1.10(6)$ using the standard methods of motional state detection~\cite{Leibfried2003}.
We believe reduction of the overlap observed here, besides the technical imperfections of the experiment, is also caused by the faster decoherence of the cat states due to dephasing and motional heating.

\begin{figure} [!ht]
\centering
   \includegraphics[width=\columnwidth]{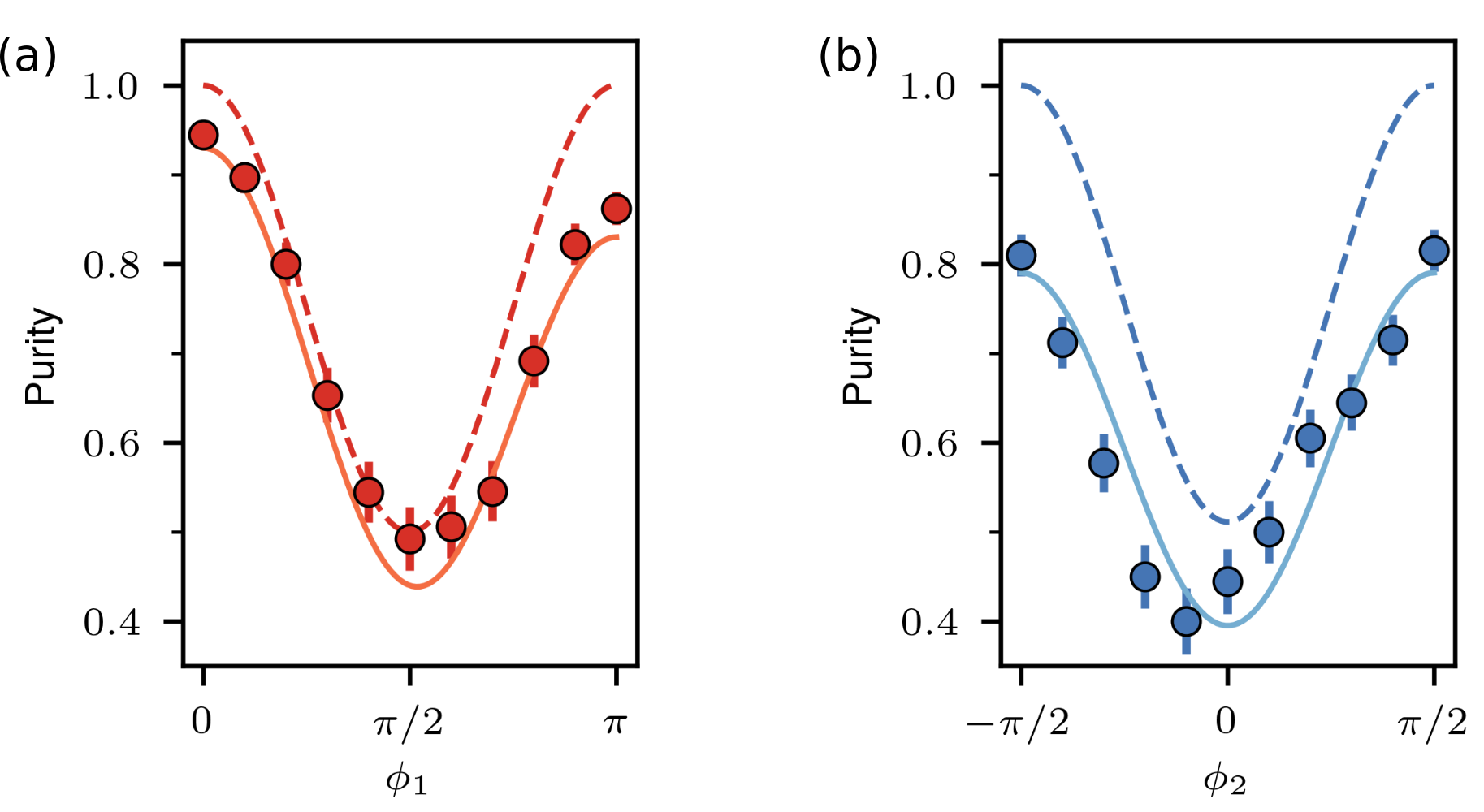}
  \caption{\label{fig:figure4} 
Measurement of purity of quantum states. (a) Measured purity of $\rho_1 =\cos^2(\frac{\phi_1}{2})|0\rangle \langle 0| + \sin^2(\frac{\phi_1}{2}) |1\rangle \langle 1 |$.
(b) Measured purity of cat states $\rho_2 = \sin^2({\frac{\phi_2}{2}+\frac{\pi}{4})|\alpha \rangle \langle \alpha |} + \cos^2(\frac{\phi_2}{2}+\frac{\pi}{4}) |-\alpha \rangle \langle -\alpha |$ where $|\alpha|^2 = 1.2(1)$. Error bars correspond to statistical uncertainty and denote one standard error of the mean.
Dash lines show expectation of purity for an ideal experiment, whereas solid lines take into account the reduction of measured overlap due to technical imperfection of the experiment (see text).
    }
  \end{figure}

An important application of SWAP test is the measurement of purity of quantum states $\rho$.
For the same density matrix $\rho$ prepared in mode $B$ and $C$, the overlap of these two modes is equivalent to the purity of $\rho$, defined as $\textrm{Tr}(\rho^2)$~\cite{ArturSwap2002}.
We demonstrate the purity measurement of quantum states that has a density matrix
$\rho = \cos^2(\nu)|\Xi\rangle \langle \Xi | + \sin^2(\nu) |\Xi_{\bot}\rangle \langle \Xi_{\bot} |$, where $|\Xi \rangle$ and $|\Xi_{\bot} \rangle$ are orthogonal. We prepare these states by entangling the internal states with the motional states, followed by optical pumping of the internal states (see Supplementary Material). 
Figure \ref{fig:figure4}a shows the measured purity of $\rho_1 = \cos^2(\frac{\phi_1}{2})|0\rangle \langle 0 | + \sin^2(\frac{\phi_1}{2}) |1\rangle \langle 1 |$ as $\phi_1$ is varied from $0$ to $\pi$.
The experimentally obtained purity is asymmetrical and lower than the expected purity determined for ideal experiments (the dashed line).
We attribute this to the fact that the overlap of Fock states measured in our experiment reduces as phonon number increases (see Fig.~\ref{fig:figure2}a).
In particular, we measured $\gamma_{11}\equiv |\langle m=1 | n=1 \rangle|^2=0.88(1)$, whereas $\gamma_{00} \equiv |\langle m=0 | n=0 \rangle|^2=0.92(1)$.
Taking this into account, the solid line in Fig.~\ref{fig:figure4}a plots the function $\gamma_{00}\cos^4(\frac{\phi_1}{2}) + \gamma_{11}\sin^4(\frac{\phi_1}{2})$, which shows good agreement with our measurement results.

In addition, we perform purity measurement for the case $\rho_2 = \sin^2({\frac{\phi_2}{2}+\frac{\pi}{4})|\alpha \rangle \langle \alpha |} + \cos^2(\frac{\phi_2}{2}+\frac{\pi}{4}) |-\alpha \rangle \langle -\alpha |$ with amplitude of coherent state $|\alpha|^2 = 1.2(1)$, which resembles a state that appears in the gedanken experiment of Schr{\"{o}}dinger~\cite{Schrodinger1935}.
By varying the parameter $\phi_2$, we drive a cat state from ``dead'' $(|-\alpha \rangle \langle -\alpha|)$ to ``alive'' $(|\alpha \rangle \langle \alpha|)$, and monitor its purity.
A ``dead or alive'' cat is described by the mixed state $\rho_2 = (|\alpha \rangle \langle \alpha| + |-\alpha \rangle \langle -\alpha| ) / 2$ obtained at $\phi_2=0$.
Figure~\ref{fig:figure4}b shows the measured purity of $\rho_2$ along with a dashed line which plots the expected purity.
The discrepancy between the measurement and the expected values can be accounted for by the imperfect overlap measurement of coherent states.
We describe this with the equation $\gamma_{\alpha}(\sin^4(\frac{\phi_2}{2}+\frac{\pi}{4})+\cos^4(\frac{\phi_2}{2}+\frac{\pi}{4}))$, shown as a solid line in Fig. \ref{fig:figure4}b. Here $\gamma_{\alpha} = 0.79(5)$ is obtained from an independent overlap measurement of coherent states with amplitude $|\alpha|^2 \sim 1$.
We observe that the measured purity of $\rho_2$ resembles that of $\rho_1$ as the purity is independent of the exact form of $|\Xi \rangle$ and $|\Xi_{\bot} \rangle$ if they are approximately orthogonal, as shown by the dashed lines in Fig.~\ref{fig:figure4}.
The purity of a $d$-dimensional quantum system satisfies $1/d \leq \textrm{Tr}(\rho^2) \leq 1$~\cite{Jaeger2006}.
The lower bound is obtained when the system is a completely mixed state.
We note that the minimum purity for both cases is close to $0.5$, which indicates that at $\phi_1=0$ and $\phi_2=0$, $\rho_1$ and $\rho_2$ are mixed states that have an effective dimension of two.

Our results may be of use in quantum information processing for states in Hilbert space with high dimension and serve as a hardware-efficient method to compute the kernels of feature vectors in quantum machine learning~\cite{Lau2017,Schuld2019}. The SWAP test presented here can be applied to other bosonic systems in which similar qubit-oscillator couplings are available, such as cavity~\cite{HarocheRMP} or circuit QED~\cite{Gao2018}, micromechanical resonators~\cite{OConnell2010}, and quantum acoustics~\cite{Chu2017}.

\begin{acknowledgments}
This research is supported by the National Research Foundation, Prime Ministers Office, Singapore, and the Ministry of Education, Singapore, under the Research Centers of Excellence program and NRF Quantum Engineering Program (Award QEP-P4).
\end{acknowledgments}
\bibliographystyle{apsrev4-1}
\bibliography{references.bib}{}
\clearpage
\widetext
\begin{center}
\textbf{\large Supplemental materials: Experimental SWAP test of infinite dimensional quantum states}
\end{center}
\setcounter{equation}{0}
\setcounter{figure}{0}
\setcounter{table}{0}
\setcounter{page}{1}
\makeatletter
\renewcommand{\theequation}{S\arabic{equation}}
\renewcommand{\thefigure}{S\arabic{figure}}
\renewcommand{\bibnumfmt}[1]{[S#1]}
\renewcommand{\citenumfont}[1]{S#1}
\section{Additional details on the SWAP test}
To measure the overlap of $|\Psi\rangle_B$ and $|\Phi\rangle_C$ in our experiment, we apply the sequence
\begin{equation*}
U^{AB}_{\textrm{CBS}}(\pi/2, \pi)U^{AC}_{\textrm{CBS}}(\pi, 0)    |g\rangle|0\rangle_A |\Psi\rangle_B |\Phi\rangle_C, 
\end{equation*} 
followed by the measurement of the state of the qubit. The probability to find the qubit in state $|g \rangle$ is denoted as $P_g$.
This SWAP test sequence is briefly explained in the main text.
Here, we give a proof that the sequence is equivalent to a controlled SWAP gate and the overlap is given as $|_{B}\langle \Psi|\Phi\rangle_{C}|^2=|1-2P_g|$.

We first describe the beam splitter transformation in more details.
A controlled beam splitter $U^{AB}_{\textrm{CBS}}(\theta,\phi)$ transforms the two creation operators corresponding to the two motional modes $A$ and $B$ as 
\begin{eqnarray}
a^{\dagger} &\rightarrow& a^{\dagger} \cos \left( \frac{\theta}{2} \right) + ib^{\dagger}e^{-i (\phi + \phi_{\textrm{s}}) } \sin \left(\frac{\theta}{2} \right) \nonumber \\ 
b^{\dagger} &\rightarrow& b^{\dagger} \cos \left( \frac{\theta}{2} \right) + i a^{\dagger}e^{i (\phi + \phi_{\textrm{s}})} \sin \left(\frac{\theta}{2} \right). \label{eqn:UCBS_trans}
\end{eqnarray}
where $\phi_{\textrm{s}}$ depends on the state of the controlled qubit. In particular, $\phi_{\textrm{s}} =0$ ($\pi$) if the qubit is $|+\rangle$ ($|-\rangle$). 
A spin-independent beam splitter $U_{\textrm{BS}}(\theta,\psi)$ has the same transformation with $\phi_\textrm{s}=0$ for all the qubit state.
Substituting $b^{\dagger}$ by $c^{\dagger}$ in the equation above gives the beam splitter transformation $U^{AC}_{\textrm{CBS}}(\theta,\phi)$ acting between mode $A$ and $C$.

For the particular case of $U^{AC}_{\textrm{CBS}}(\pi, 0)$, we have
\begin{eqnarray}
a^{\dagger} &\rightarrow& ie^{-i\phi_\textrm{s}}c^{\dagger} \nonumber \\
c^{\dagger} &\rightarrow& ie^{i\phi_\textrm{s}}a^{\dagger}. \label{eqn:1stcbs}
\end{eqnarray}

For $U^{AB}_{\textrm{CBS}}(\pi/2, \pi)$,
\begin{eqnarray}
a^{\dagger} &\rightarrow& \frac{1}{\sqrt{2}}(a^\dagger -ie^{-i\phi_\textrm{s}}b^\dagger) \nonumber\\
b^{\dagger} &\rightarrow& \frac{1}{\sqrt{2}}(b^\dagger -ie^{i\phi_\textrm{s}}a^\dagger) \label{eqn:2ndcbs}.
\end{eqnarray}

We begin the proof by considering the case $|\Psi \rangle_B = |m\rangle_B$ and $|\Phi\rangle_C = |n\rangle_C$.
The initial state can be expressed as
\begin{equation*}
    |g\rangle|0\rangle_A |m\rangle_B |n\rangle_C = 
    \frac{1}{\sqrt{2m! n!}}(b^\dagger)^{m}(c^\dagger)^{n}(|+\rangle -|-\rangle )|\textrm{vac}\rangle,
\end{equation*}
where $|\textrm{vac}\rangle$ denotes $|0\rangle_A |0 \rangle_B |0 \rangle_C $. To simplify the notation, we will ignore the normalization factor $\frac{1}{\sqrt{m! n!}}$ and label the state $|g\rangle|0\rangle_A |m\rangle_B |n\rangle_C$ as $|g,0,m,n\rangle$.

Applying the first controlled beam splitter to the initial state results in
\begin{equation*}
    U^{AC}_{\textrm{CBS}}(\pi, 0)    |g,0,m,n\rangle = \frac{1}{\sqrt{2}}\left[ (b^{\dagger})^m (ia^\dagger)^{n} |+,\textrm{vac}\rangle - (b^{\dagger})^m (-ia^\dagger)^{n} |-,\textrm{vac}\rangle\right],
\end{equation*}
which is obtained by using the relation \ref{eqn:1stcbs}.
Using the relation \ref{eqn:2ndcbs}, the second controlled beam splitter transforms the state into the final state $|\Theta_{\textrm{fin}}\rangle$, given by
\begin{equation}
    |\Theta_{\textrm{fin}}\rangle = \frac{1}{\sqrt{2^{m+n+1}}}(b^{\dagger}-ia^{\dagger})^{m} (b^{\dagger}+ia^{\dagger})^{n}|+,\textrm{vac}\rangle -
    \frac{1}{\sqrt{2^{m+n+1}}}(b^{\dagger}+ia^{\dagger})^{m} (b^{\dagger}-ia^{\dagger})^{n}|-,\textrm{vac}\rangle. \label{eqn:finstate}
\end{equation}
The probability of measuring the qubit in $|g \rangle$ is then given as $P_g=\textrm{Tr}_{\textrm{ext}}(|g\rangle \langle g| \rho_\textrm{fin})$ where $\rho_\textrm{fin}=|\Theta_{\textrm{fin}}\rangle \langle \Theta_{\textrm{fin}}|$ and $\textrm{Tr}_{\textrm{ext}}$ denotes the partial trace over all motional modes.

To simplify the calculation of $P_g$, we apply two spin-independent controlled beam splitters $U^{AC}_{\textrm{BS}}(\pi,\pi)$ and $U^{AB}_{\textrm{BS}}(\pi/2,0)$ to the final state 
\begin{equation*}
    |\Theta'\rangle = U^{AC}_{\textrm{BS}}(\pi,\pi) U^{AB}_{\textrm{BS}}(\pi/2,0)|\Theta_{\textrm{fin}}\rangle.
\end{equation*}
The probability to detect the qubit in $|g\rangle$ of $|\Theta'\rangle$ is given as $P'_g=\textrm{Tr}(|g\rangle \langle g| \rho')$ where $\rho'=|\Theta'\rangle \langle \Theta'|$, and can be related to $\rho_\textrm{fin}$ as
\begin{equation*}
    P'_g = \textrm{Tr}(|g\rangle \langle g |U_{\textrm{m}} \rho_\textrm{fin}U^{\dagger}_{\textrm{m}}),
\end{equation*}
where $U_{\textrm{m}}$ denotes $U^{AC}_{\textrm{BS}}(\pi,\pi) U^{AB}_{\textrm{BS}}(\pi/2,0)$.

Using the cyclic property of the trace operator and the fact that the unitary operator $U_\textrm{m}$ acts on only the motional modes and hence commutes with $|g\rangle \langle g|$, we can rewrite $P'_g$ as
\begin{equation*}
\textrm{Tr}(U^{\dagger}_{\textrm{m}}|g\rangle \langle g |U_{\textrm{m}} \rho_\textrm{fin})  =  \textrm{Tr}(U^{\dagger}_{\textrm{m}}U_{\textrm{m}}|g\rangle \langle g| \rho_\textrm{fin}) = \textrm{Tr}(|g\rangle \langle g| \rho_\textrm{fin}) ,
\end{equation*}
which is equal to $P_g$. Therefore, applying spin-independent beam splitters after our gate sequence does not alter the measurement results of the qubit and we can obtain the expression for $P_g$ by evaluating $P'_g$. To do this, we first need to derive $|\Theta'\rangle$.

Using \ref{eqn:UCBS_trans} with $\phi_\textrm{s}=0$, $\theta=\pi/2$, and $\phi=0$, the transformation of the corresponding creation operators due to $U^{AB}_{\textrm{BS}}(\pi/2,0)$ is
\begin{eqnarray*}
    \frac{1}{\sqrt{2}}(b^{\dagger} - ia^{\dagger}) &\rightarrow& b^\dagger\\
    \frac{1}{\sqrt{2}}(b^{\dagger} + ia^{\dagger}) &\rightarrow& i a^\dagger.
\end{eqnarray*}
Substituting these relations into \ref{eqn:finstate}, we have
\begin{equation*}
    U^{AB}_{\textrm{BS}}(\pi/2,0)|\Theta_{\textrm{fin}}\rangle =\frac{1}{\sqrt{2}}(b^\dagger)^m(ia^\dagger)^n |+,\textrm{vac}\rangle - \frac{1}{\sqrt{2}}(ia^\dagger)^m(b^\dagger)^n|-,\textrm{vac}\rangle.
\end{equation*}
The last beam splitter $U^{AC}_{\textrm{BS}}(\pi,\pi)$ has the following transformation
\begin{eqnarray*}
    a^\dagger &\rightarrow& -ic^\dagger\\
    c^\dagger &\rightarrow& - ia^\dagger.
\end{eqnarray*}
Using this transformation, we arrive at
\begin{equation*}
    |\Theta'\rangle=U^{AC}_{\textrm{BS}}(\pi,\pi) U^{AB}_{\textrm{BS}}(\pi/2,0) |\Theta_{\textrm{fin}}\rangle = \frac{1}{\sqrt{2}}(b^\dagger)^m (c^\dagger)^n |+,\textrm{vac}\rangle -\frac{1}{\sqrt{2}}(c^\dagger)^m (b^\dagger)^n |-,\textrm{vac}\rangle,
\end{equation*}
which can be rewritten as
\begin{equation}
    |\Theta'\rangle=\frac{1}{\sqrt{2}}|+\rangle|0, m, n \rangle - \frac{1}{\sqrt{2}} |-\rangle|0, n, m \rangle. \label{eqn:cswap}
\end{equation}
The motional mode $A$ is returned to the vacuum state and the motional mode $B$ and $C$ are swapped if the qubit is $|-\rangle$.
This shows that our SWAP test sequence is equivalent to a controlled  SWAP gate.

Next we consider a general case in which $|\Psi\rangle_B = \sum^{d}_{m=0} \beta_{m}|m\rangle$ and $|\Phi\rangle_C = \sum^{d}_{n=0} \zeta_{n}|n\rangle$ where $d$ denotes the dimension of the quantum states encoded in the motional modes $B$ and $C$.
The above sequence transforms this initial state into
\begin{equation*}
    |\Theta'\rangle = \frac{1}{2}\sum_{m,n} \beta_m \zeta_{n}\Big[ |+\rangle|0, m,n \rangle -  |-\rangle|0,n,m \rangle\Big].
\end{equation*}
Substituting this into $P'_g = \textrm{Tr}_{\textrm{ext}}(|g\rangle \langle g|\Theta'\rangle \langle \Theta'|)$, we can evaluate $P'_g$ as
\begin{equation*}
\frac{1}{4}\textrm{Tr}_\textrm{ext} \Big[ \sum_{m,n,l,k} \beta_m \zeta_n \beta_l^{*} \zeta_k^{*} \Big(|g,0,m,n\rangle \langle g,0,l,k| - |g,0,m,n\rangle \langle g,0,k,l| -|g,0,n,m\rangle \langle g,0,l,k| + |g,0,n,m\rangle \langle g,0,k,l| \Big) \Big]
\end{equation*}
Noting that $\textrm{Tr}_\textrm{ext}(|g,0,m,n\rangle \langle g,0,l,k|) = \delta_{ml}\delta_{nk}$ where $\delta_{ml}$ and $\delta_{nk}$ are the Kronecker delta symbols, we obtain
\begin{eqnarray*}
    P'_g &=&  \frac{1}{2}\sum_{m,n,l,k} \beta_m \zeta_n \beta_l^{*} \zeta_k^{*} \Big[\delta_{ml}\delta_{nk}-\delta_{mk}\delta_{nl}\Big]\\
    &=& \frac{1}{2}\sum_{m,n}|\beta_m|^2|\zeta_n|^2 -\frac{1}{2}|\sum_{k}\beta_k \zeta^{*}_k|^2.
\end{eqnarray*}
Using  $\sum_{m}|\beta_m|^2 =1 $, $\sum_{n}|\zeta_n|^2=1$, and $|\langle \Psi | \Phi \rangle|^2= |\sum_{k}\beta_k \zeta^{*}_k|^2$, we arrive at
$|\langle \Psi | \Phi \rangle|^2 = |1-2P'_g|=|1-2P_g|$.
\section{State preparation for purity measurement}
In the experiment, we measure the purity of $\rho_1 = \cos^2(\frac{\phi_1}{2})|0\rangle \langle 0 | + \sin^2(\frac{\phi_1}{2}) |1\rangle \langle 1 |$ and $\rho_2 = \sin^2({\frac{\phi_2}{2}+\frac{\pi}{4})|\alpha \rangle \langle \alpha |} + \cos^2(\frac{\phi_2}{2}+\frac{\pi}{4}) |-\alpha \rangle \langle -\alpha |$ where $|\alpha\rangle$ is the coherent state and $|\alpha|^2=1.2(1)$.
Here, we describe the state preparation procedure of these states.

We prepare $\rho_1$ by applying the sequence
\begin{equation*}
R(\theta_1) U^{B}_{\textrm{bsb}}(\phi_1 )|g\rangle |0\rangle_B
\end{equation*}
to mode $B$, followed by 5\,$\mu$s of optical pumping.
The operation $ U^{B}_{\textrm{bsb}}(\phi_1 )$ represents the blue motional sideband pulse applied on mode $B$ with a pulse area of $\phi_1$.
After applying $ U^{B}_{\textrm{bsb}}(\phi_1 )$, the state is
\begin{equation*}
U^{B}_{\textrm{bsb}}(\phi_1 )|g\rangle |0\rangle_B=\cos \left(\frac{\phi_1}{2} \right)|g\rangle |0\rangle_B + \sin \left( \frac{\phi_1}{2} \right )|e\rangle |1\rangle_B.
\end{equation*}
The rotation $R(\theta_1)$ is applied to keep the excited state population of the qubit always below 0.5 before the optical pumping.
This reduces heating of motional modes due to the photon recoil. The rotation angle is set as $\theta_1 = 0$ when $ 0 \leq \phi_1 \leq \pi/2$, resulting in the state
\begin{equation*}
|\Psi\rangle_1=\cos \left(\frac{\phi_1}{2} \right)|g\rangle |0\rangle_B + \sin \left( \frac{\phi_1}{2} \right )|e\rangle |1\rangle_B,
\end{equation*}
and $\theta_1= \pi$ for $ \pi/2 < \phi_1 < \pi$, 
\begin{equation*}
|\Psi\rangle_1=\cos \left(\frac{\phi_1}{2} \right)|e\rangle |0\rangle_B + \sin \left( \frac{\phi_1}{2} \right )|g\rangle |1\rangle_B.
\end{equation*}
Optical pumping resets the qubit, reducing the pure entangled state between the qubit and motion to a mixed state of the motional mode which is described by $\rho_1 = \textrm{Tr}_\textrm{qubit}(|\Psi\rangle_1 \langle \Psi|_1)= \cos^2(\frac{\phi_1}{2})|0\rangle \langle 0 | + \sin^2(\frac{\phi_1}{2}) |1\rangle \langle 1 |$.
 The same sequence is repeated on mode $C$.
 
 We prepare  $\rho_2 = \sin^2({\frac{\phi_2}{2}+\frac{\pi}{4})|\alpha \rangle \langle \alpha |} + \cos^2(\frac{\phi_2}{2}+\frac{\pi}{4}) |-\alpha \rangle \langle -\alpha |$  by applying the sequence 
 \begin{equation}
R(\theta_2)D^{B}_{x}(\alpha) R(\phi_2) |g\rangle |0\rangle_B,
\end{equation}
followed by 5\,$\mu$s of optical pumping. Here, $D^{B}_{x}(\alpha) \equiv e^{\sigma_x(\alpha b^{\dagger} - \alpha^{*}b)}$ represents a coherent displacement of mode $B$ by an amount $\alpha$, conditioned on the qubit states and transforms the states as follows,
\begin{eqnarray}
    D^{B}_{x}(\alpha)|+\rangle |0\rangle_B &=& | + \rangle |\alpha \rangle_B \\
    D^{B}_{x}(\alpha)|-\rangle |0\rangle_B &=& | - \rangle |-\alpha \rangle_B.
\end{eqnarray}
where $\alpha$ is a real number.
Using the above transformation, we have
\begin{eqnarray*}
    D^{B}_{x}(\alpha) R(\phi_2) |g\rangle |0\rangle_B &=& \frac{1} {\sqrt{2}} \left[ \cos\left( \frac{\phi_2}{2} \right) +\sin\left( \frac{\phi_2}{2} \right) \right]| + \rangle |\alpha \rangle_B + \frac{1} {\sqrt{2}} \left[ \sin\left( \frac{\phi_2}{2} \right) - \cos\left( \frac{\phi_2}{2} \right) \right]  | - \rangle |-\alpha \rangle_B\\
    &=& \sin \left( \frac{\phi_2}{2} + \frac{\pi}{4} \right )| + \rangle |\alpha \rangle_B - \cos \left( \frac{\phi_2}{2} + \frac{\pi}{4} \right )| - \rangle |-\alpha \rangle_B
\end{eqnarray*}
The rotation $R(\theta_2)$ serves a similar role to reduce the excited state population of the qubit before the optical pumping. 
For $ -\pi/2 \leq \phi_2 < 0$, we set $\theta_2= \pi/2$ and obtain the state
\begin{equation*}
    |\Psi\rangle_2=\sin \left( \frac{\phi_2}{2} + \frac{\pi}{4} \right )| e \rangle |\alpha \rangle_B - \cos \left( \frac{\phi_2}{2} + \frac{\pi}{4} \right )| g \rangle |-\alpha \rangle_B.
\end{equation*}
For $ 0 \leq \phi_2 \leq \pi/2$, we set $\theta_2 = - \pi/2$ and the resulting state is
\begin{equation*}
    |\Psi\rangle_2=\sin \left( \frac{\phi_2}{2} + \frac{\pi}{4} \right )| g \rangle |\alpha \rangle_B - \cos \left( \frac{\phi_2}{2} + \frac{\pi}{4} \right )| e \rangle |-\alpha \rangle_B.
\end{equation*}
After optical pumping which results in tracing out the qubit, we obtain the final state $\rho_2 = \textrm{tr}_\textrm{qubit}(|\Psi\rangle_2 \langle \Psi|_2) = \sin^2({\frac{\phi_2}{2}+\frac{\pi}{4})|\alpha \rangle \langle \alpha |} + \cos^2(\frac{\phi_2}{2}+\frac{\pi}{4}) |-\alpha \rangle \langle -\alpha |$.
We apply the same sequence to prepare $\rho_2$ on mode $C$.
\end{document}